# Proximity-effect-induced Superconducting Gap in Topological Surface States – A Point Contact Spectroscopy Study of $NbSe_2/Bi_2Se_3$ Superconductor-Topological Insulator Heterostructures

Wenqing Dai[1], Anthony Richardella[1], Renzhong Du[1], Weiwei Zhao[1], Xin Liu[1], C.X. Liu[1], Song-Hsun Huang[2], Raman Sankar[2], Fangcheng Chou[2], Nitin Samarth[1], and Qi Li[*1]

[1]*Department of Physics, The Pennsylvania State University, University Park, Pennsylvania 16802, USA*

[2]*Center for Condensed Matter Sciences, National Taiwan University, Taipei 10617, Taiwan*

Proximity-effect-induced superconductivity was studied in epitaxial topological insulator $Bi_2Se_3$ thin films grown on superconducting $NbSe_2$ single crystals. A point contact spectroscopy (PCS) method was used at low temperatures down to 40 mK. An induced superconducting gap in $Bi_2Se_3$ was observed in the spectra, which decreased with increasing $Bi_2Se_3$ layer thickness, consistent with the proximity effect in the bulk states of $Bi_2Se_3$ induced by $NbSe_2$. At very low temperatures, an extra point contact feature which may correspond to a second energy gap appeared in the spectrum. For a 16 quintuple layer $Bi_2Se_3$ on $NbSe_2$ sample, the bulk state gap value near the top surface is ~ 159 μeV, while the second gap value is ~ 120 μeV at 40 mK. The second gap value decreased with increasing $Bi_2Se_3$ layer thickness, but the ratio between the second gap and the bulk state gap remained about the same for different $Bi_2Se_3$ thicknesses. It is plausible that this is due to superconductivity in $Bi_2Se_3$ topological surface states induced through the bulk states. The two induced gaps in the PCS measurement are consistent with the three-dimensional bulk state and the two-dimensional surface state superconducting gaps observed in the angle-resolved photoemission spectroscopy (ARPES) measurement.

[*] Correspondence and requests for materials should be addressed to Q.L. (email: qil1@psu.edu).

## Introduction

Since the first experimentally accessible proposal of topological superconductors (TSCs) by Fu and Kane,[1] the search for TSCs and Majorana zero modes has generated significant interest in condensed matter physics. Majorana zero modes exist at the boundary of TSCs and have potential applications in quantum computing. TSCs may exist intrinsically in superconducting doped topological insulators (TI),[2-9] however there is a lack of experimental evidence of this due to the lack of intrinsic TSCs. Alternatively, TSCs can also occur in proximity-induced superconductors, such as in a TI in contact with an s-wave superconductor.[10-22] For this reason, significant efforts have been devoted to proximity-induced topological superconductivity. Electron tunneling and point contact spectroscopy (PCS), which probe the density of states (DOS) of superconductors, have been widely used in studies of topological superconductor systems and in the search for Majorana fermions. Some studies reported zero-bias conductance peak (ZBCP) features in the transport spectra of S/TI junctions and point contact spectra on S/TI bilayers,[15,23-27] which were proposed as a signature of TSCs. On the other hand, scanning tunneling spectroscopy (STS) measurements on epitaxial TI thin films on superconducting $NbSe_2$ substrates showed no ZBCP.[13,19] In the core of a magnetic vortex inside a topological insulator-superconductor bilayer, a ZBCP was observed by STS and attributed to Majorana fermions.[21,22] Recently, an angle-resolved photoemission spectroscopy (ARPES) study revealed two dimensional topological superconductivity in proximity coupled $NbSe_2/Bi_2Se_3$ heterostructures.[20] The topological nature of this superconducting state was demonstrated by the observation of spin-momentum locking. Since there is no clear conclusive observation of the topological superconducting states in transport studies, it is very important to know how the proximity-induced topological superconducting states in the same $NbSe_2/Bi_2Se_3$ samples will be reflected in the transport measurements.

In this work, we present point contact spectroscopy studies of epitaxial $Bi_2Se_3$ thin films with different thicknesses grown on $NbSe_2$ single crystals at low temperatures down to 40 mK. The samples were grown in identical conditions as those measured by ARPES.[20] While the ARPES measurements are only sensitive to the surface layer of the order of nm, the PCS measurements are sensitive to the depth within the proximity coherence length range or the mean free path whichever is smaller, which is around 16 nm for the bulk $Bi_2Se_3$. Our results showed that a finite superconducting energy gap was successfully induced in the $Bi_2Se_3$ from the superconducting $NbSe_2$ substrate through the proximity effect. No ZBCP features were present in the point contact spectra. The induced gap decreased with increasing $Bi_2Se_3$ thickness. More importantly, a second gap-like feature appeared besides the main gap at temperatures below 0.45 K in the point contact spectra of a16 QL $Bi_2Se_3$ on $NbSe_2$ sample, suggesting a possible signature of an induced superconducting gap in topological surface states through the bulk states. The second gap also decreased with increasing $Bi_2Se_3$ thickness, but the ratio between the second gap and the bulk state gap remained about the same for different $Bi_2Se_3$ thicknesses. The two induced gaps from our point contact spectroscopy measurements are consistent with the ARPES results that superconducting gaps are induced by proximity effect in both the $Bi_2Se_3$ bulk states and the surface states.

## Results

**Soft point contact technique.** We studied the point contact spectroscopy of $NbSe_2$ single crystal and $NbSe_2/Bi_2Se_3$ heterostructure samples by using the "soft" point contact technique.[28] The contacts were made by applying a small drop of silver paint between the top surface of the sample and a gold wire of 25 μm diameter (Fig. 1). The other electrical contacts were made on the $NbSe_2$ superconducting substrate for four-probe electrical measurements, as show schematically in the inset of Fig. 1. Unlike conventional Needle-Anvil point contacts, silver paint "soft" point contacts utilize nanometer-scale Ag particles to form parallel contact channels to the samples in which there is no pressure applied to the sample. Thus it is preferred in studying properties on the surfaces in thin and soft $Bi_2Se_3$ films.

Depending on the ratio of the electron mean free path to the radius of contact channels, point contacts can be categorized into ballistic, diffusive, or thermal regimes.[28] It is very important to verify that the point contact is in the ballistic regime. Non-ideal features, such as dips at voltage values larger than the superconducting gap, can appear in the conductance curves when the contact is not ballistic. If the radius of a point contact is much larger than the mean free path, the contact is in the thermal regime and no spectroscopic information can be obtained. In addition, the thermal contact region temperature is higher than the bath temperature from local Joule heating. We carefully examined our point contact spectra to ensure all data reported were from ballistic contacts.

**Proximity-effect-induced gap in $Bi_2Se_3$.** The conductance spectra of point contacts on samples of different $Bi_2Se_3$ film thicknesses at low temperatures are shown in Fig. 2**a**. For samples with thin $Bi_2Se_3$ layers (<7 QL), the $NbSe_2$ gap feature (~ 1.0 mV) dominates the spectrum. With increasing $Bi_2Se_3$ film thickness, another conductance peak feature appears at low voltage bias (~ 0.3 mV) and becomes more and more pronounced. Eventually, only this peak feature is present in the 16 QL sample. For the samples with multiple peaks in the spectra, we fitted the experimental data using an extended Blonder-Tinkham-Klapwijk (BTK) model[29] which assumes a linear combination of two different gaps and independent fitting parameters $\Delta$, $\Gamma$, and Z for each gap. The BTK theory is widely used to describe the transport between a normal metal and a superconductor with a finite transparency of the interface. The parameter $\Gamma$ was included to describe the broadening effect, which is associated with the lifetime of the quasiparticles.[30] The fitted gap values $\Delta$ from PCS measurements of all samples are plotted in Fig. 2**b** together with $Bi_2Se_3$ bulk gaps at the top surface from the ARPES measurement.[20] The PCS gaps are clearly separated into two groups. The $NbSe_2$ gap value decreases slightly in the $Bi_2Se_3/NbSe_2$ heterostructures from the gap of pure $NbSe_2$ (~ 1.2 meV) with the increasing $Bi_2Se_3$ layer thickness. Meanwhile, a smaller gap feature appears in the PCS spectra when the $Bi_2Se_3$ film thickness is above 10 QL and follows the same thickness dependence trend as the induced $Bi_2Se_3$ bulk band gaps from ARPES measurements on the same type of $NbSe_2/Bi_2Se_3$ heterostructures. Therefore, this gap feature is the proximity-induced bulk state gap in $Bi_2Se_3$ near the surface. The $NbSe_2$ gap and induced bulk state gap show very different magnetic field dependences. The conductance spectrum peaks from the induced gap ( ~ 0.3 mV) are quickly suppressed in small magnetic fields. On the other hand, the $NbSe_2$ gap feature at ~ 1.0 mV disappears under ~ 4 T magnetic field, which is consistent with the $H_{c2}$ of $NbSe_2$ (See Supplementary Information A and B for more details).

Unlike ARPES, which is only sensitive to a few monolayers from the top surface, PCS probes the top surface with a deeper depth into the sample with a length of the order of the mean free path $l_e$ in dirty limit.[28] The $Bi_2Se_3$ film has an electron density of $n \sim 1.3\times10^{19} cm^{-3}$ and in-plane residual resistivity $\rho_0^{ab} \sim 0.75 m\Omega\cdot cm$.[31] The Fermi level of $Bi_2Se_3$ films is in the conduction band, about ~ 0.4 eV above the Dirac point.[20] Based on the resistivity anisotropy measurements of single crystals and selected orientation thin films made by MBE,[32,33] we take the resistivity anisotropic ratio to be $\frac{\rho_c}{\rho_{ab}}$ ~ 4. Using a c-direction effective mass $m_c^* = 0.76 m_e$ from the results of reflectance studies of $Bi_2Se_3$ crystals[34] and $v_F^c$ $= 2.39\times10^5 m/s$ from the band structure calculations[20], the out-of-plane mean free path $l_e^c = \frac{m_c^* v_F^c}{\rho_c n e^2}$ is estimated to be ~ 16 nm for the bulk $Bi_2Se_3$. This is very close to the $Bi_2Se_3$ thickness threshold when no $NbSe_2$ gap is observed. When the $Bi_2Se_3$ film thickness is much smaller than $l_e$, the signal is mainly from the interface of $NbSe_2$ substrate. With increasing $Bi_2Se_3$ film thickness, the gap from the $Bi_2Se_3$ film starts to gain more weight in the total spectra. Eventually, when the $Bi_2Se_3$ film thickness is beyond $l_e$, the gap in $Bi_2Se_3$ dominates the spectra and the signal from the interface vanishes. Therefore, this also supports that the gap values from 0.30 to 0.16 meV in 10 to 16 QL samples in Fig. 2**b** are the proximity-effect-induced superconducting energy gap in the bulk states of the $Bi_2Se_3$ thin film. It should be noted that the above spectra are obtained on samples with medium contact transparency so that the spectrum is close to tunneling regime. When the point contact is very clean (low barrier strength), we observed Andreev reflection spectra which were from the $NbSe_2/Bi_2Se_3$ interface, as shown in Fig. 5**a** of ref. 20. We did not observe ZBCP features in the spectra, consistent with STS measurement results[13,19], but different from some PCS measurement results[15,23-25]. Our results, similar to the STS results, also call for a careful re-examination of the interpretation of ZBCP in the PCS as the signature of unconventional superconductivity or Majorana fermions.

Now we focus on the samples with only proximity-effect-induced gap feature. Figure 3**a** shows the conductance spectra of a point contact junction on a $NbSe_2$/16 QL $Bi_2Se_3$ sample at low temperatures down to 40 mK. Only the proximity-induced gap is present in the spectra, without a signal from the $NbSe_2$ substrate contribution. Spectra at temperatures from 1.8 K to 7.5 K are plotted in Fig. 3**b**. As the temperature increases, the induced gap peaks in the conductance spectra start to smear into a single broad peak at zero bias likely due to the thermal broadening. The zero bias conductance then decreases gradually till 7 K, $T_c$ of $NbSe_2$, confirming that this gap is due to the proximity effect from the superconductor $NbSe_2$. Figure 3**c** plots the conductance spectra to high voltage bias at temperatures up to 15 K. The spectra shows a linear background, which is often attributed to inelastic tunneling at the point contact–sample interface.[35-38] The conductance spectra at ~ 60 mK under different magnetic fields are plotted in Fig. 3**d**. The gap feature is suppressed by a field less than 0.3 T, which is much smaller than the $H_{c2}$ of $NbSe_2$, ~ 4 T. It is likely due to that the broadening from the pair-breaking effect in magnetic fields smears the $Bi_2Se_3$ bulk gap feature in the conductance spectra (see Supplementary Information C for discussion).

**Additional gap-like feature at low temperatures.** We fitted the conductance spectra with the BTK model. Figure 4**a** plots the normalized conductance spectra of the point contact junction on the $NbSe_2$/16 QL $Bi_2Se_3$ at low temperatures down to 40 mK. The barrier strength parameter Z from the fittings is ~ 1 for all temperatures, indicating high tunneling barrier strength. The fitted gap value $\Delta$ and broadening parameter Γ versus temperature $T$ are shown in Fig. 4**b**. We find that at very low temperatures below 0.45 K, the fitting curves using the standard BTK model deviate from the experimental conductance spectra as shown in Fig. 4**a**. If we force to use a single gap fitting, the gap value from the fittings shows an abnormal decrease at low temperatures (circled in Fig. 4**b**), which cannot be true as the superconductivity and the proximity-coupling enhance with decreasing temperatures. The conductance curves can be fitted nicely using the BTK model above 0.45 K, indicating the point contact is in the ballistic regime and the anomaly at low temperature is not from diffusive or thermal contacts. Using the same Δ, Γ, and Z parameters as in the last good fitting curve at 0.45 K, a conductance curve is calculated for $T$ = 40 mK and plotted together with the experimental data in Fig. 5. While the outside edges of peaks agree well for the two curves, the inner gap edge is narrower for the experimental data. The calculated curve is then subtracted from the experimental data and the difference is plotted in the top right inset of Fig. 5. The excess spectrum shows a clear peak feature at ~ 120 μV. The peak position remains unchanged as the temperature is increased from 40 mK to 260 mK. Figure 6**a** plots the low magnetic field dependence of the PCS spectra on the $NbSe_2$/16 QL $Bi_2Se_3$ heterostructure at $T$ ~ 43 mK. Deviations of fittings using the BTK model from experimental data are also visible below 0.02 T. In PCS studies under magnetic field, Γ is often used to simulate the pair-breaking effect of a magnetic field in a first-order approximation.[28] We fitted the PCS data under magnetic field at both 60 mK and 1.8 K with the BTK model and observed a Γ/$H$ ratio of ~3 meV/T at low magnetic field for both temperatures (Supplementary Information C). Using the Γ/$H$ ratio and the BTK parameters listed in Fig. 5 and assuming that the main gap at 159 μeV doesn't change in small magnetic fields, we employed a similar method to calculate the conductance difference between the experimental data and simulated curves. The results are plotted in Fig. 6**b**. Although the peak position does not change much, the peak magnitude at ~ 120 μV is suppressed by a small magnetic field ~ 0.03 T. From the temperature dependence and magnetic field dependence of the PCS data, an additional gap feature ~ 120 μeV at 40 mK seems to appear besides the induced bulk band gap ~ 159 μeV. This second gap feature is not observable in the spectra above 0.45 K, likely because that the thermal broadening ($kT$ ~ 86 μeV/K) at high temperatures smears the difference between the two gaps (the two gap energy difference ~ 39 μeV at 40 mK) in the point contact spectra. Similarly, the second gap feature in the spectra is smeared in a small magnetic field ~ 0.03 T likely due to the broadening effect from the pair-breaking in magnetic fields. Spectroscopic measurement with much higher energy resolution is desired to resolve more accurately at what temperature and the magnetic field the second gap disappears. A similar second gap feature is also present in a 13 QL $Bi_2Se_3$ sample at low temperatures (see Supplementary Information D). The second gap feature is ~ 210 μeV with a bulk gap ~ 0.3 meV at 0.4 K. The ratio of the second gap value to the bulk gap value is ~ 0.7, close to the two gap ratio in the16 QL $Bi_2Se_3$ sample at similar temperatures. This indicates that the second gap feature in 16 QL and 13 QL samples are from the same origin. The second gap value decreases with increasing $Bi_2Se_3$ thickness, but the ratio between the second gap and the bulk state gap stays constant, independent of the $Bi_2Se_3$ thickness.

## Discussions

There are several possible origins of multiple proximity-effect-induced gap values. First, in "soft" point contact measurement, the silver paint contact area is much larger than the real electrical contact size, parallel conductance channels may form in the contact area. In this case, a second gap measured from another contact will be present in the spectra. However, in our data, the second gap feature is observed in both 13 QL and 16 QL samples with a ratio ~ 0.7 to the main induced gap at 0.2 to 0.4 K. Therefore, it is not likely from the parallel contact channels to a non-uniform sample surface. Secondly, $NbSe_2$ is a multiband superconductor. High resolution scanning tunneling microscopy (STM) measurements on pure $NbSe_2$ single crystal reveals a main tunneling peak at 1.2 meV with a shoulder at 0.75 meV.[39-42] In our data, only one gap ~ 1.2 meV is visible down to low temperatures in the PCS spectra on pure $NbSe_2$ and $NbSe_2/Bi_2Se_3$ heterostructure with very thin $Bi_2Se_3$ (see Fig. S1**b** in supplementary information). Therefore, the second gap feature is not from the multiple superconducting gaps in $NbSe_2$. Thirdly, proximity effect is greatly affected by the boundary conditions [43]. In the case of S/N bilayer samples, multiple Andreev bound states may exist when the N layer thickness is larger than the proximity coherence length.[44] In our samples, the $Bi_2Se_3$ film thickness is much smaller than the $Bi_2Se_3$ proximity coherence length $\xi_N^c = \sqrt{\frac{\hbar v_F^c l_e^c}{6\pi k_B T}} = \frac{40nm}{\sqrt{T}}$ at low temperatures. Therefore, the two induced gap features are not from the multiple bound states in proximity coupled $Bi_2Se_3$ bulk states.

In the previous ARPES studies on the $NbSe_2/Bi_2Se_3$ samples,[20] S.Y. Xu *et al.* showed that both the $Bi_2Se_3$ bulk states and the topological surface states become superconducting due to the proximity effect from the superconducting $NbSe_2$. Here our result on the same $NbSe_2/Bi_2Se_3$ heterostructures suggests that there is clearly a proximity-induced bulk state gap, but there is also a second gap at low temperatures which could be the topological surface state gap as observed in the ARPES measurements. While the gap ~ 159 μeV in the16 QL $Bi_2Se_3$ sample is from the superconducting bulk states, the gap feature ~ 120 μeV at 40 mK could be a signature of the proximity-induced superconductivity in the $Bi_2Se_3$ surface states on the top surface. In the ARPES measurements, it was reported that the surface-to-surface (interface) hybridization may enhance the helical pairing in the surface states on the top surface when the $Bi_2Se_3$ layer thickness is below 6 QL. As $Bi_2Se_3$ thickness increases ($\geq$ 6 QL), the surface state wave function from the top surface and the interface become spatially separated. Therefore, in thick $Bi_2Se_3$ samples (13 QL and 16 QL), $NbSe_2$ does not directly induce superconductivity in the top surface states through surface hybridization; the superconducting surface state gap at top surface is induced through the bulk bands of $Bi_2Se_3$. In our results, The ratio of the superconducting surface state gap to bulk state gap value is ~ 0.7 in both 13 QL and 16 QL samples between 0.2 K to 0.4 K, indicating a similar coupling strength from the bulk to the surface bands in different thickness samples. The second gap feature is only observed at ultra-low temperatures for all samples. This is largely due to that the energy gap of the surface states is very close to the bulk gap. Decreasing temperature improves the spectrum energy resolution by reducing the thermal broadening ($kT$ ~ 86 μeV/K) in the PCS measurements. Our results suggest that cooling to ultra-low temperature is desired to resolve the superconducting surface states from the superconducting bulk states in electron spectroscopic measurements, which is consistent with other reports that the signatures from the TI surface states were only observed at very low temperatures.[15,16]

The ARPES measurements showed that the induced superconductivity in the surface states has a unique spin-momentum locking and Dirac-dispersion nature, different from the induced superconductivity in the bulk states. The standard BTK model employed in our PCS data fitting is based on the s-wave superconductivity. While it could describe the induced superconducting gap in the bulk states satisfactorily, it does not apply to the induced gap in the topological surface states due to its unique two dimensional helical-Cooper pairing phase. Indeed, we were not able to fit the point contact spectra of the 16 QL sample below 0.45 K using an extended BTK model with a linear combination of two different gaps. A model developed for helical pairing is desired to characterize the superconductivity in the topological surface states in the future.

In summary, we conducted point contact spectroscopy measurements on $NbSe_2/Bi_2Se_3$ heterostructures down to 40 mK. We observed a proximity-effect-induced superconductivity gap in the bulk of the TI $Bi_2Se_3$ thin film, which is consistent with the bulk state gap values reported by the ARPES measurements on similar samples. The induced bulk gap value is ~ 159 μeV at 0.45 K for a 16 QL $Bi_2Se_3$ on $NbSe_2$ sample. Below 0.45 K, excess conductance spectra appeared which may correspond to a second gap feature. This could be due to the topological superconductivity gap as observed by ARPES. The induced second gap value is ~ 120 μeV at 40 mK for a 16 QL $Bi_2Se_3$ sample. The second gap spectra peaks were suppressed at temperatures above 0.45 K or in a magnetic field of ~ 0.03 T, likely due to thermal or field-induced broadening effect in the spectra. The second gap feature was also observed in a $NbSe_2$/13 QL $Bi_2Se_3$ sample with a similar two gap value ratio ~ 0.7 at around 0.4 K. As the ARPES measurements confirmed both the $Bi_2Se_3$ bulk state and topological surface state become superconducting from the proximity effect, it is plausible that the second induced gap feature in the point contact spectra is due to the induced superconductivity in the topological surface states. The PCS result suggests that ultra-low temperature is desirable to separate the superconducting surface states from the bulk states in electron spectroscopic measurements.

## Methods

Single phase epitaxial $Bi_2Se_3/NbSe_2$ heterostructure samples were fabricated using molecular beam epitaxy (MBE) method, which has been described in detail elsewhere.[20] The contacts were made by applying a small drop of silver paint between the top surface of the sample and a gold wire of 25 μm diameter. Immediately after point contacts were made on fresh $NbSe_2/Bi_2Se_3$ heterostructures, samples were cooled in a Quantum Design Physical Property Measurement System (PPMS) for four probe electrical measurements at temperatures down to 1.8 K. A Quantum Design dilution refrigerator system was used for measurements down to 40 mK. The differential conductance spectra were obtained by a lock-in technique in which a 262 Hz AC modulation of less than 50 μV was applied to the sample in addition to a DC current bias. Mathematical derivatives of the *I-V* data were also used to confirm the measurement, but with lower resolutions. The AC modulation amplitude was reduced to be less than $\frac{1}{2}k_B T$ for measurements below 1 K.

## Data Availability

The data generated during the current study are available from the corresponding author on request.

## Acknowledgements

We thank Moses Chan, Jainendra Jain, Su-Yang Xu, Zahid Hasan and Fan Zhang for helpful discussions. The authors would like to acknowledge the support from United States Department of Energy under Grant No. DE-FG02-08ER46531 (Q.L.) and DE-SC0005042 (J.J) and from the Ministry of Science and Technology in Taiwan under Grant No. MOST-104-2119-M-002 -028-MY2 (F.C.). This study is based upon research conducted at The Pennsylvania State University Two-Dimensional Crystal Consortium – Materials Innovation Platform (2DCC-MIP) which is supported by NSF cooperative agreement DMR-1539916.

## Author contributions

W.D. conducted point contact measurements and model fittings of the data; R.D. and W.Z. took part in low temperature measurements; A.R. and N.S. grew the $Bi_2Se_3$ films by MBE; S-H.H., R.S., and F.C. prepared the $NbSe_2$ single crystals; X.L. and C.X.L. performed theoretical calculations; Q.L. was responsible for the design, planning, and supervision of the overall experimental work. W.D. was responsible for drafting the manuscript and Q.L., A.R., and N.S. were mainly responsible for the revision of the manuscript. All authors contributed to the revision of the manuscript.

## Competing financial interests

The authors declare no competing financial interests.

## Reference

[1] Fu, L. & Kane, C. Superconducting Proximity Effect and Majorana Fermions at the Surface of a Topological Insulator. *Phys. Rev. Lett.* **100,** 1–4 (2008).

[2] Hor, Y.S. *et al.* Superconductivity in $Cu_xBi_2Se_3$ and its Implications for Pairing in the Undoped Topological Insulator. *Phys. Rev. Lett.* **104,** 3–6 (2010).

[3] Fu, L. & Berg, E. Odd-Parity Topological Superconductors: Theory and Application to $Cu_xBi_2Se_3$. *Phys. Rev. Lett.* **105,** 97001 (2010).

[4] Sasaki, S. *et al.* Topological Superconductivity in $Cu_xBi_2Se_3$. *Phys. Rev. Lett.* **107,** 3–7 (2011).

[5] Hsieh, T. & Fu, L. Majorana Fermions and Exotic Surface Andreev Bound States in Topological Superconductors: Application to $Cu_xBi_2Se_3$. *Phys. Rev. Lett.* **108,** 1–5 (2012).

[6] Levy, N. *et al.* Experimental Evidence for s-Wave Pairing Symmetry in Superconducting $Cu_xBi_2Se_3$ Single Crystals Using a Scanning Tunneling Microscope. *Phys. Rev. Lett.* **110,** 117001 (2013).

[7] Peng, H., De, D., Lv, B., Wei, F. & Chu, C.-W. Absence of zero-energy surface bound states in $Cu_xBi_2Se_3$ studied via Andreev reflection spectroscopy. *Phys. Rev. B* **88,** 24515 (2013).

[8] Yonezawa, S. *et al.* Thermodynamic evidence for nematic superconductivity in $Cu_xBi_2Se_3$. *Nat. Phys.* **13,** 123–126 (2016).

[9] Liu, Z. *et al.* Superconductivity with Topological Surface State in $Sr_xBi_2Se_3$. *J. Am. Chem. Soc.* ***137*** (33), 10512–10515 (2015).

[10] Zhang, D. *et al.* Superconducting proximity effect and possible evidence for Pearl vortices in a candidate topological insulator. *Phys. Rev. B* **84,** 165120 (2011).

[11] Qu, F. *et al.* Strong Superconducting Proximity Effect in Pb-$Bi_2Te_3$ Hybrid Structures. *Sci. Rep.* **2,** 339 (2012).

[12] Veldhorst, M. *et al.* Josephson supercurrent through a topological insulator surface state. *Nat. Mater.* **11,** 1–5 (2012).

[13] Wang, M.-X. *et al.* The coexistence of superconductivity and topological order in the $Bi_2Se_3$ thin films. *Science* **336,** 52–5 (2012).

[14] Williams, J. *et al.* Unconventional Josephson Effect in Hybrid Superconductor-Topological Insulator Devices. *Phys. Rev. Lett.* **109,** 1–5 (2012).

[15] Yang, F. *et al.* Proximity effect at superconducting Sn-$Bi_2Se_3$ interface. *Phys. Rev. B* **85,** 104508 (2012).

[16] Yang, F. *et al.* Proximity-effect-induced superconducting phase in the topological insulator $Bi_2Se_3$. *Phys. Rev. B* **86,** 134504 (2012).

[17] Zareapour, P. *et al.* Proximity-induced high-temperature superconductivity in the topological insulators $Bi_2Se_3$ and $Bi_2Te_3$. *Nat. Commun.* **3,** 1056 (2012).

[18] Wang, E. *et al.* Fully gapped topological surface states in Bi2Se3 films induced by a d-wave high-temperature superconductor. *Nat. Phys.* **9,** 621–625 (2013).

[19] Xu, J.-P. *et al.* Artificial Topological Superconductor by the Proximity Effect. *Phys. Rev. Lett.* **112,** 217001 (2014).

[20] Xu, S.-Y. *et al.* Momentum-space imaging of Cooper pairing in a half-Dirac-gas topological superconductor. *Nat. Phys.* **10,** 943–950 (2014).

[21] Xu, J.-P. *et al.* Experimental Detection of a Majorana Mode in the core of a Magnetic Vortex inside a Topological Insulator-Superconductor $Bi_2Te_3/NbSe_2$ Heterostructure. *Phys. Rev. Lett.* **114,** 17001 (2015).

[22] Sun, H. H. *et al*. Majorana Zero Mode Detected with Spin Selective Andreev Reflection in the Vortex of a Topological Superconductor. *Phys. Rev. Lett.* **116**, 1–5 (2016).

[23] Koren, G., Kirzhner, T., Kalcheim, Y. & Millo, O. Signature of proximity-induced $p_x + ip_y$ triplet pairing in the doped topological insulator $Bi_2Se_3$ by the s-wave superconductor NbN. *Europhysics Lett.* **103,** 67010 (2013).

[24] Koren, G., Kirzhner, T., Lahoud, E., Chashka, K.B. & Kanigel, A. Proximity-induced superconductivity in topological $Bi_2Te_2Se$ and $Bi_2Se_3$ films: Robust zero-energy bound state possibly due to Majorana fermions. *Phys. Rev. B* **84,** 224521 (2011).

[25] Koren, G. & Kirzhner, T. Zero-energy bound states in tunneling conductance spectra at the interface of an s-wave superconductor and a topological insulator in NbN/$Bi_2Se_3$/Au thin-film junctions. *Phys. Rev. B* **86,** 144508 (2012).

[26] Koren, G. Proximity effects at the interface of a superconductor and a topological insulator in NbN-$Bi_2Se_3$ thin film bilayers. *Supercond. Sci. Technol.* **28**, 025003 (2015).

[27] Li, H. *et al.* The origin of bias independent conductance plateaus and zero bias conductance peaks in $Bi_2Se_3$/$NbSe_2$ hybrid structures. *arXiv*:1607.07731 (2016).

[28] Daghero, D. & Gonnelli, R.S. Probing multiband superconductivity by point-contact spectroscopy. *Supercond. Sci. Technol.* **23,** 43001 (2010).

[29] Blonder, G.E., Tinkham, M. & Klapwijk, T.M. Transition from metallic to tunneling regimes in superconducting microconstrictions: Excess current, charge imbalance, and supercurrent conversion. *Phys. Rev. B* **25,** 4515 (1982).

[30] Dynes, R.C., Narayanamurti V., & Garno, J.P. Direct Measurement of Quasiparticle-Lifetime Broadening in a Strong-Coupled Superconductor. *Phys. Rev. Lett.* **41**, 1509 (1978).

[31] Kandala, A., Richardella, A., Zhang, D., Flanagan, T.C. & Samarth, N. Surface-Sensitive Two-Dimensional Magneto-Fingerprint in Mesoscopic $Bi_2Se_3$ Channels. *Nano Lett.* **13,** 2471–6 (2013).

[32] Stordeur, M., Ketavong, K.K., Priemuth, A., Sobotta, H. & Riede, V. Optical and Electrical Investigations of n-Type $Bi_2Se_3$ Single Crystals. *Phys. Stat. Sol. (b)* **169,** 505–514 (1992).

[33] Xu, Z. *et al.* Anisotropic Topological Surface States on High-Index $Bi_2Se_3$ Films. *Adv. Mater.* **25,** 1557–62 (2013).

[34] Gobrecht, H. & Seeck, S. Effective Mass Dependence on Carrier Concentration in $Bi_2Se_3$. *Z. Physik.* **222,** 93–104 (1969).

[35] Kirtley, J.R. & Scalapino, D.J. Inelastic-tunneling model for the linear conductance background in the high-Tc superconductors. *Phys. Rev. Lett.* **65**, 798 (1990).

[36] Kirtley, J.R., Washburn, S. & Scalapino, D.J. Origin of the linear tunneling conductance background. *Phys. Rev. B* **45**, 336 (1992).

[37] Kirtley, J.R. Inelastic transport through normal-metal–superconductor interfaces. *Phys. Rev. B* **47**, 11379 (1993).

[38] Chen, X., Huan, C., Hor, Y.S., Sa de Melo, C.A.R., & Jiang, Z. Point-contact Andreev reflection spectroscopy of candidate topological superconductor $Cu_{0.25}Bi_2Se_3$. *arXiv*: 1210.6054 (2012).

[39] Noat, Y. *et al.* Signatures of multigap superconductivity in tunneling spectroscopy. *Phys. Rev. B* **82,** 014531 (2010).

[40] Guillamon, I., Suderow, H., Guinea, F. & Vieira, S. Intrinsic atomic-scale modulations of the superconducting gap of 2*H*-$NbSe_2$. *Phys. Rev. B* **77,** 134505 (2008).

[41] Rodrigo, J.G. & Vieira, S. STM study of multiband superconductivity in $NbSe_2$ using a superconducting tip. *Phys. C Supercond.* **404,** 306–310 (2004).

[42] Fridman, I., Kloc, C. & Wei, J.Y.T. Cryomagnetic scanning tunneling spectroscopy study of multi-gap spectra in superconducting 2*H*-$NbSe_2$. *arXiv*:1110.6490 (2011).

[43] Kim, J. *et al.* Visualization of geometric influences on proximity effects in heterogeneous superconductor thin films. *Nat. Phys.* **8,** 465–470 (2012).

[44] Truscott, A.D., Dynes, R.C. & Schneemeyer, L.F. Scanning Tunneling Spectroscopy of $NbSe_2$-Au Proximity Junctions. *Phys. Rev. Lett.* **83,** 1014–1017 (1999).

## Figure Legends

**Figure 1. Point contacts on $NbSe_2/Bi_2Se_3$ heterostructure.** An optical microscopic picture showing two silver paint point contacts on a $NbSe_2/Bi_2Se_3$ heterostructure sample. The Au wire diameter is 25 um. The size of the silver paint contacts is ~ 50 um. A schematic of the point contact spectroscopy measurement of the bilayer sample is shown in the inset.

**Figure 2. Point contact spectra of $NbSe_2/Bi_2Se_3$ heterostructures of different $Bi_2Se_3$ thicknesses. a,** Normalized conductance spectra of point contacts on $NbSe_2/Bi_2Se_3$ samples of different $Bi_2Se_3$ thicknesses at ~ 0.1 K. The $R_n$ values are 485 Ω, 71 Ω, 26 Ω, and 57 Ω, respectively. Curves are shifted vertically for clarity. **b,** Energy gaps from point contact measurements plotted together with bulk state gaps from ARPES measurements on the $NbSe_2/Bi_2Se_3$ heterostructures of various $Bi_2Se_3$ film thicknesses. The red line is a guide for the eye.

**Figure 3. Point contact spectra of a $NbSe_2$/16 QL $Bi_2Se_3$ heterostructure. a,** from 0.04 K to 7.0 K. **b**, from 1.8 K to 7.5 K. **c,** Conductance spectra to high *V* bias at temperatures 1.8 K, 5.0 K, 7.5 K, 10 K, and 15 K, respectively. **d,** Conductance spectra at different magnetic fields. The sample temperature is at ~ 60 mK. The contact normal resistance $R_n$ is 57 Ω.

**Figure 4. Fittings to point contact spectra of a $NbSe_2$/16 QL $Bi_2Se_3$ heterostructure at low temperatures. a,** Normalized conductance spectra of a point contact on a $NbSe_2$/16 QL $Bi_2Se_3$ sample at low temperatures. Curves are shifted vertically for clarity. The black lines are BTK model fits to the experimental data. **b,** The temperature dependence of the superconducting energy gap Δ and broadening parameter Γ values from the BTK model fittings in Fig. 4a.

**Figure 5. Additional gap-like feature in point contact spectrum of the $NbSe_2$/16 QL $Bi_2Se_3$ heterostructure at 40 mK.** A point contact spectrum of the $NbSe_2$/16 QL $Bi_2Se_3$ heterostructure at 40 mK (black) plotted together with a BTK model simulated curve using the parameters shown in the bottom left of the figure (red). The top right inset shows the conductance difference between the experimental data and the simulation as a function of bias *V*. The dash lines mark positions of peaks at ~ 120 μV.

**Figure 6. Magnetic field dependence of the additional gap-like feature. a,** Point contact spectra of the $NbSe_2$/16 QL $Bi_2Se_3$ heterostructure at low magnetic fields. The sample temperature is ~ 43 mK. Curves are shifted vertically for clarity. The black lines are BTK model fits to the experimental data. **b,** The conductance difference between the experimental data and simulations vs. bias *V* at different magnetic fields. Curves are shifted vertically for clarity.

# Figures

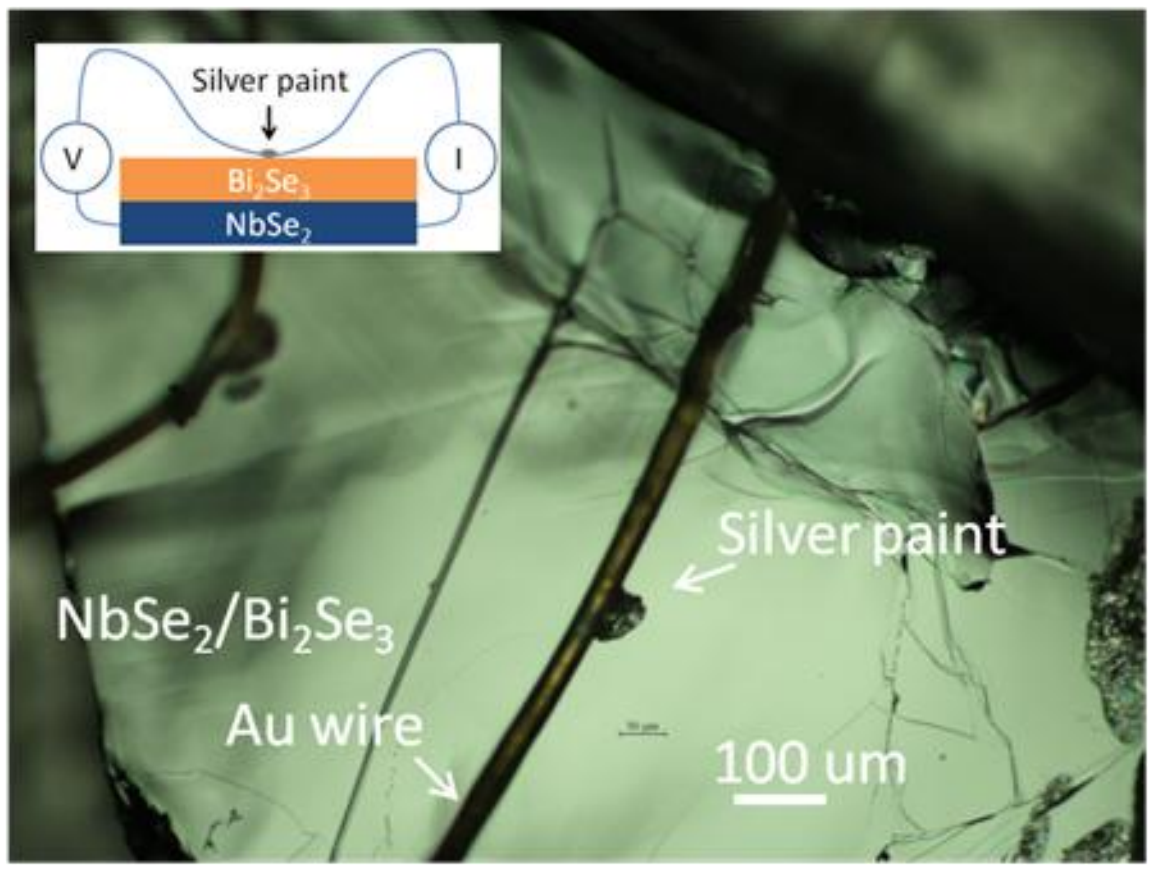

**Figure 1**

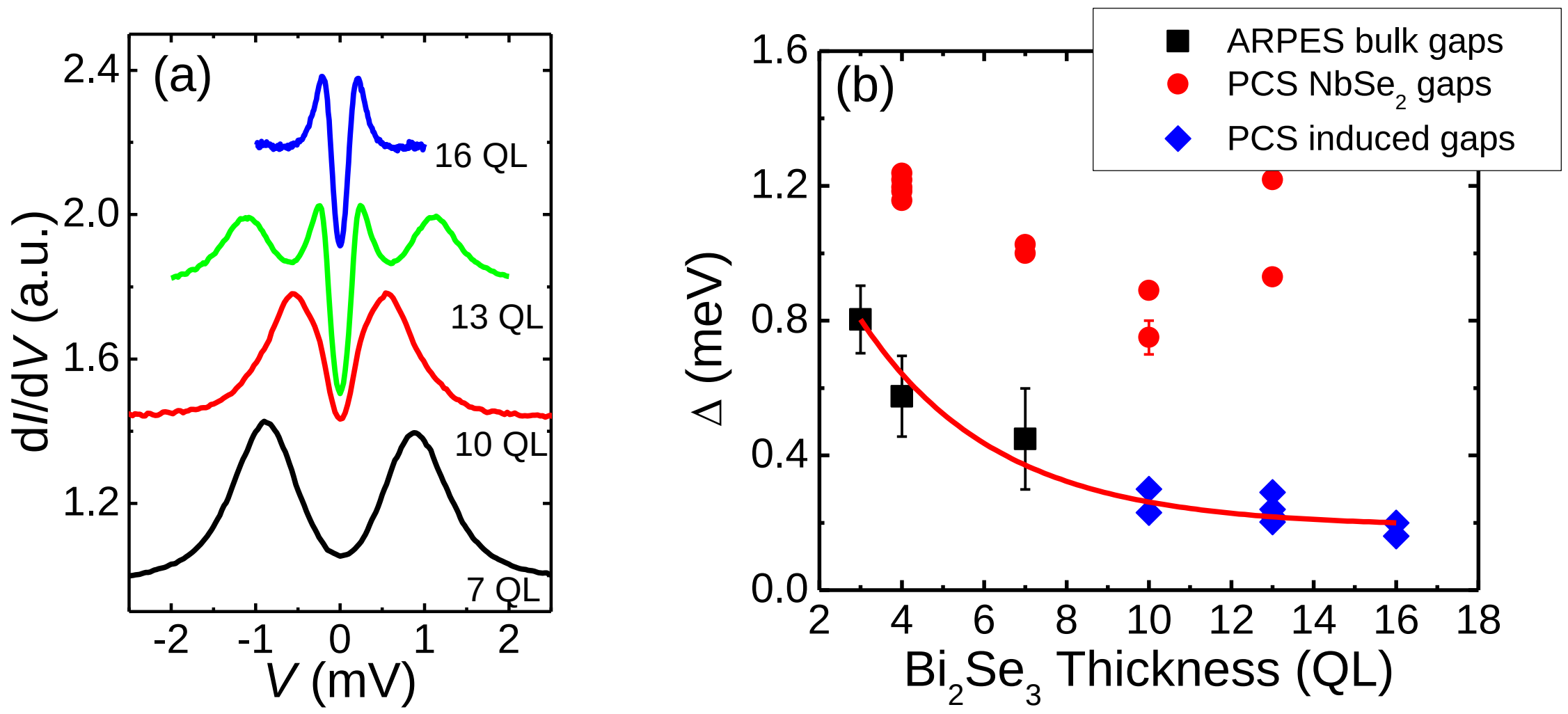

**Figure 2**

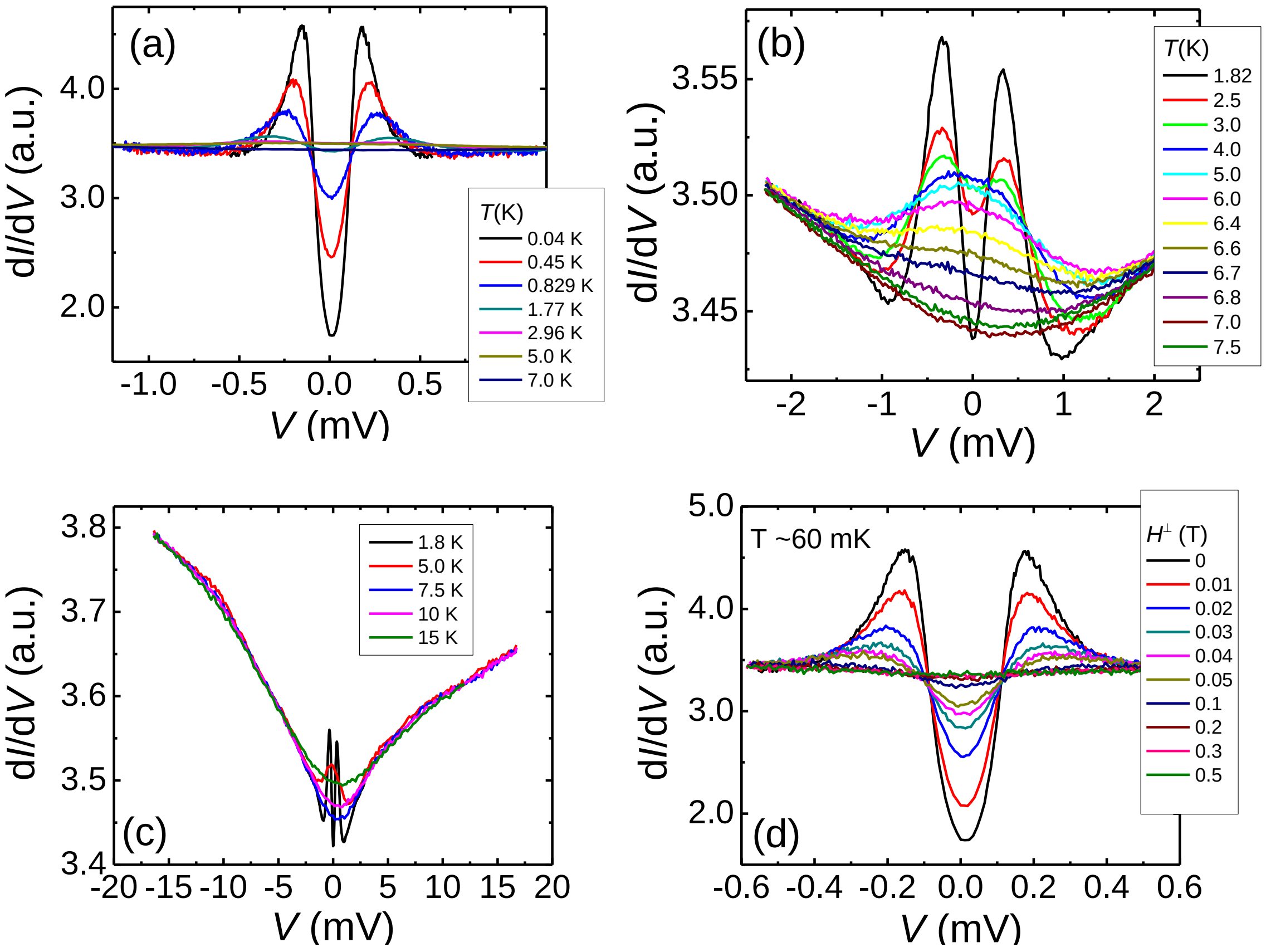

**Figure 3**

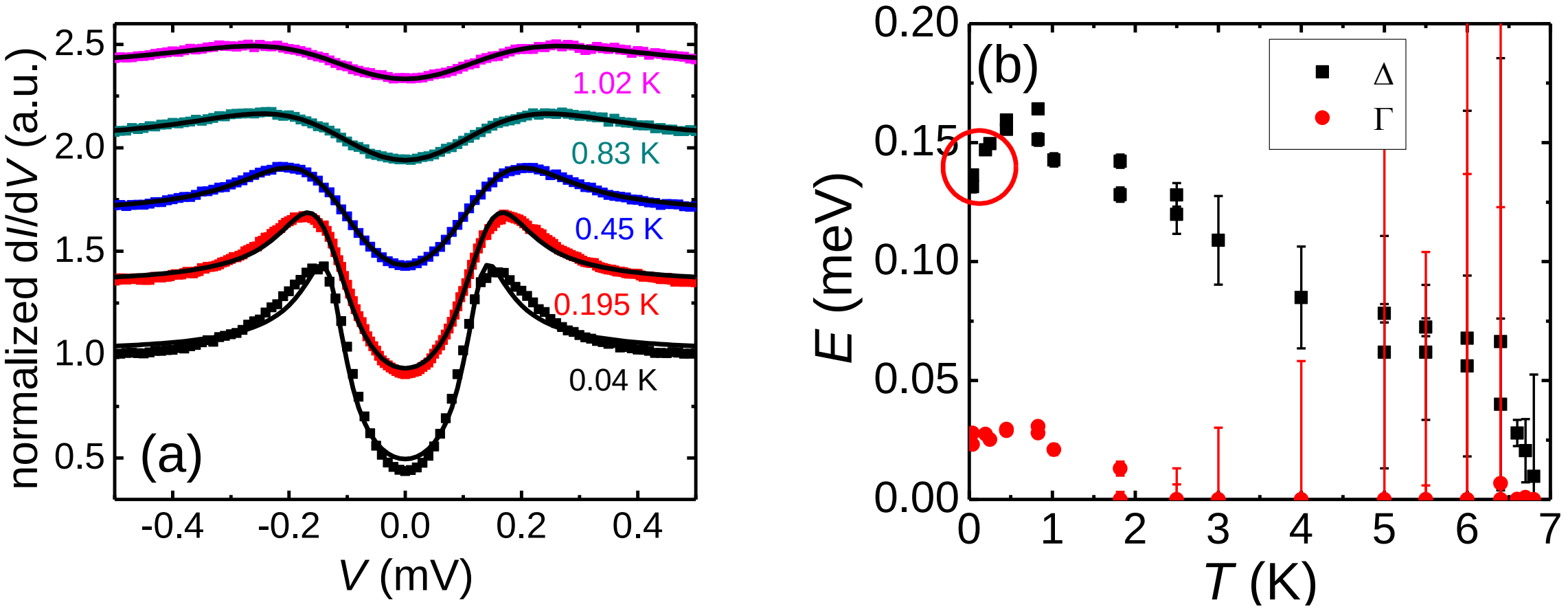

**Figure 4**

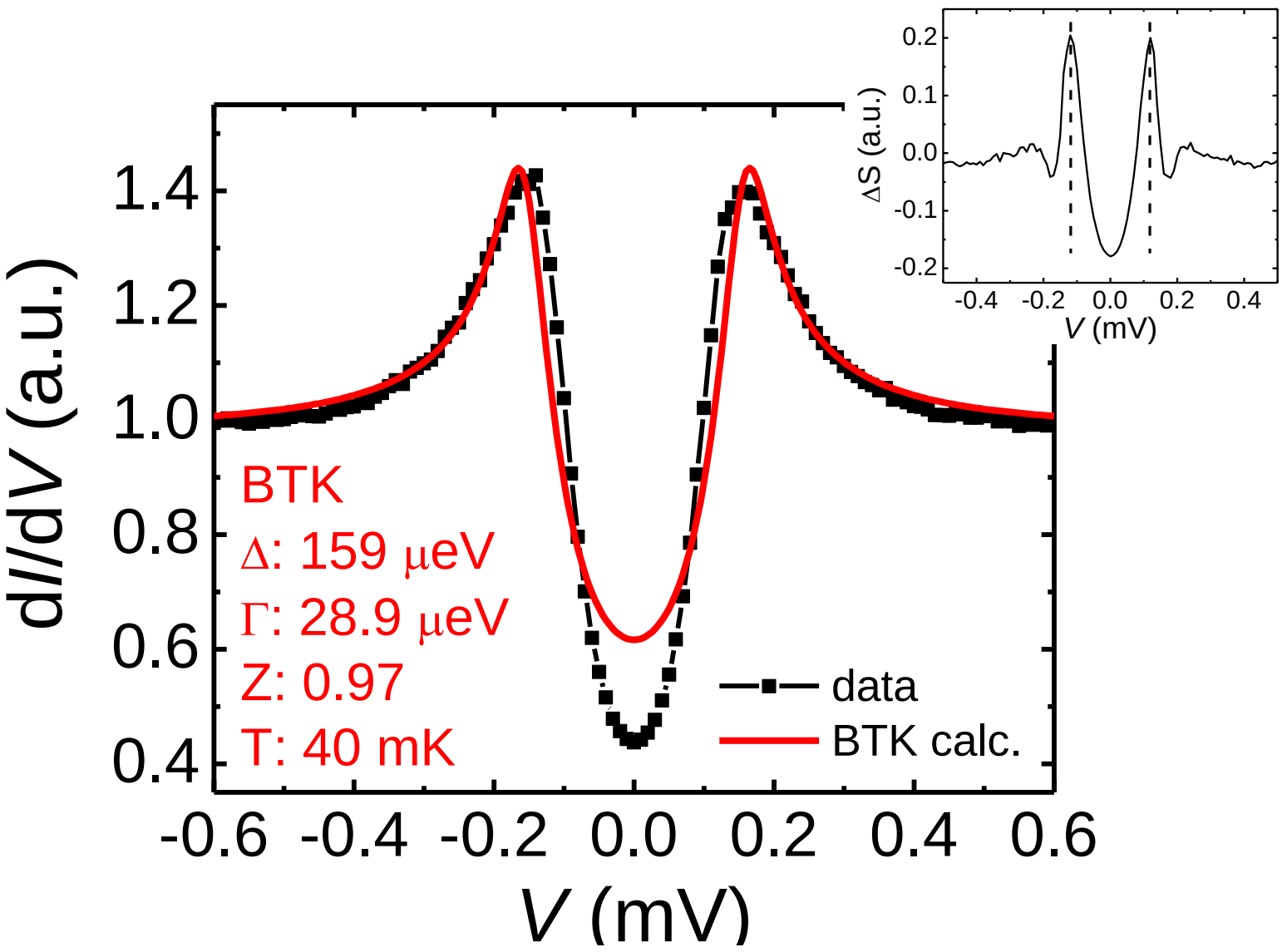

**Figure 5**

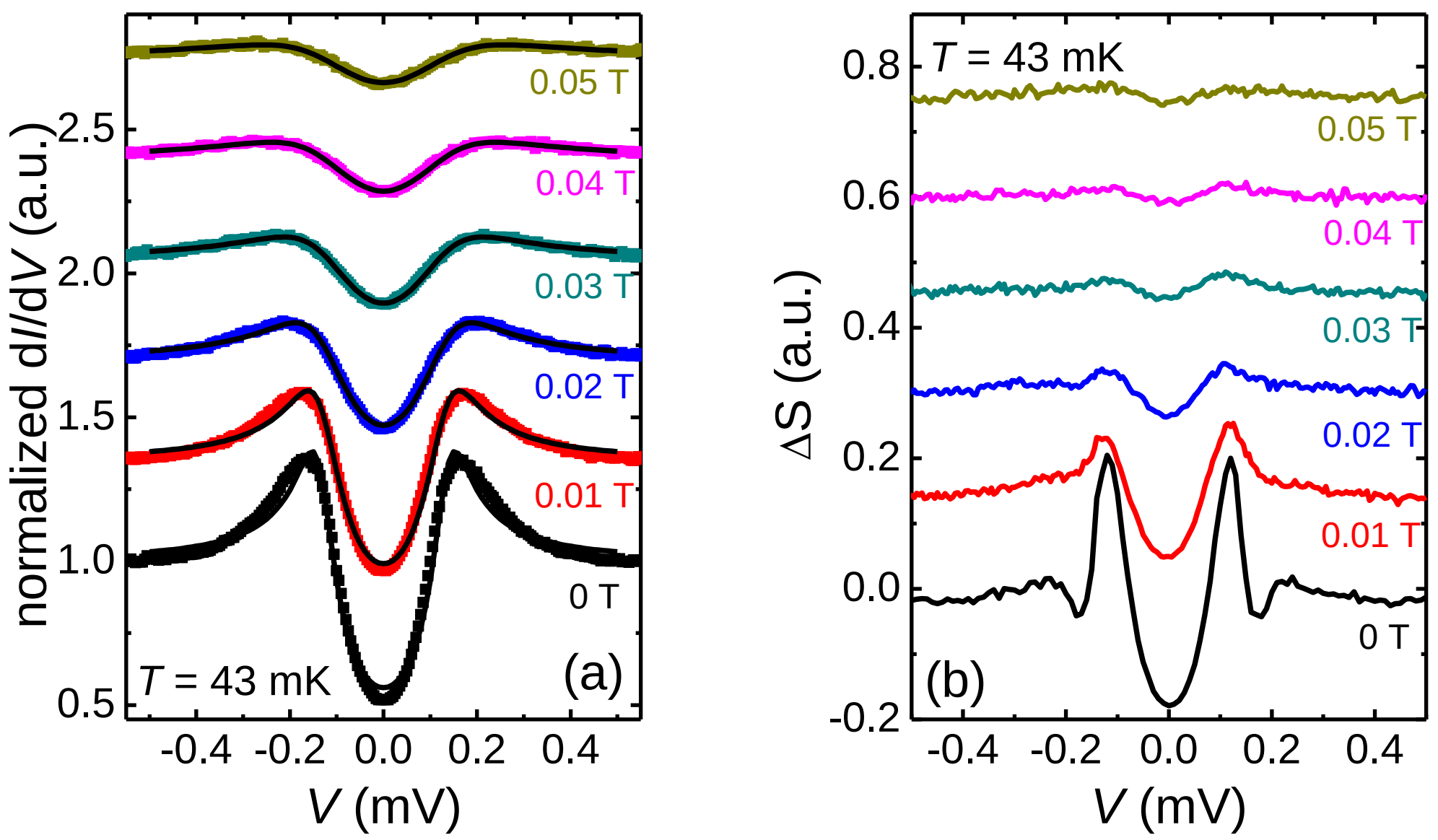

**Figure 6**

# Supplementary Information

# Proximity-effect-induced Superconducting Gap in Topological Surface States – A Point Contact Spectroscopy Study of $NbSe_2/Bi_2Se_3$ Superconductor-Topological Insulator Heterostructures

Wenqing Dai, Anthony Richardella, Renzhong Du, Weiwei Zhao, Xin Liu, C.X. Liu, Song-Hsun Huang,

Raman Sankar, Fangcheng Chou, Nitin Samarth, and Qi Li

This file includes

## SI A. Point contact spectra of a $NbSe_2$/7 QL $Bi_2Se_3$ heterostructure

We have measured point contact spectra of pure $NbSe_2$ single crystal and very thin $Bi_2Se_3$ films on $NbSe_2$. The pure $NbSe_2$ result was presented in supplementary information E of Ref. 1. Figure S1**a** plots the conductance spectra at different temperatures of a point contact on a $NbSe_2$/7 QL $Bi_2Se_3$ heterostructure and the fittings using the Blonder-Tinkham-Klapwijk (BTK) theory,[2] which is widely used to describe the transport between a normal metal and a superconductor with a finite transparency of the interface. A parameter $\Gamma$ was included to describe the broadening effect.[3] The spectra were fitted well with the BTK theory and the energy gap value $\Delta$ and broadening parameter $\Gamma$ from the fittings are plotted in Fig. S1**b**. Figure S1**c** shows the conductance spectra in different magnetic fields at 0.2 K and the $\Delta$ and $\Gamma$ values from the fittings to the spectra are plotted in Fig. S1**d**. Here $\Gamma$ is used, as a first approximation, to simulate the pair-breaking effect of a magnetic field.[3] The fitted energy gap $\Delta$ is ~ 1.0 meV at the lowest temperature, which is slightly lower than the pure $NbSe_2$ gap value ~ 1.26 meV from point contact measurement.[1] The gap value decreases with increasing temperature or magnetic field. The energy gap feature in the spectra disappears at ~ 7 K or under ~ 4 T magnetic field, which is consistent with the $T_c$ and $H_{c2}$ of $NbSe_2$ single crystals.

The measured superconducting energy gap in Fig. S1**b** and S1**d** are from the interface of the $NbSe_2$/$Bi_2Se_3$ which is dominated by the superconducting gap of $NbSe_2$ at the interface. This gap value decreases slightly from the bulk $NbSe_2$ gap and also decreases when the thickness of $Bi_2Se_3$ films is increased. These are consistent with the suppression of the energy gap in a superconductor at the interface with another material due to proximity effect.[4]

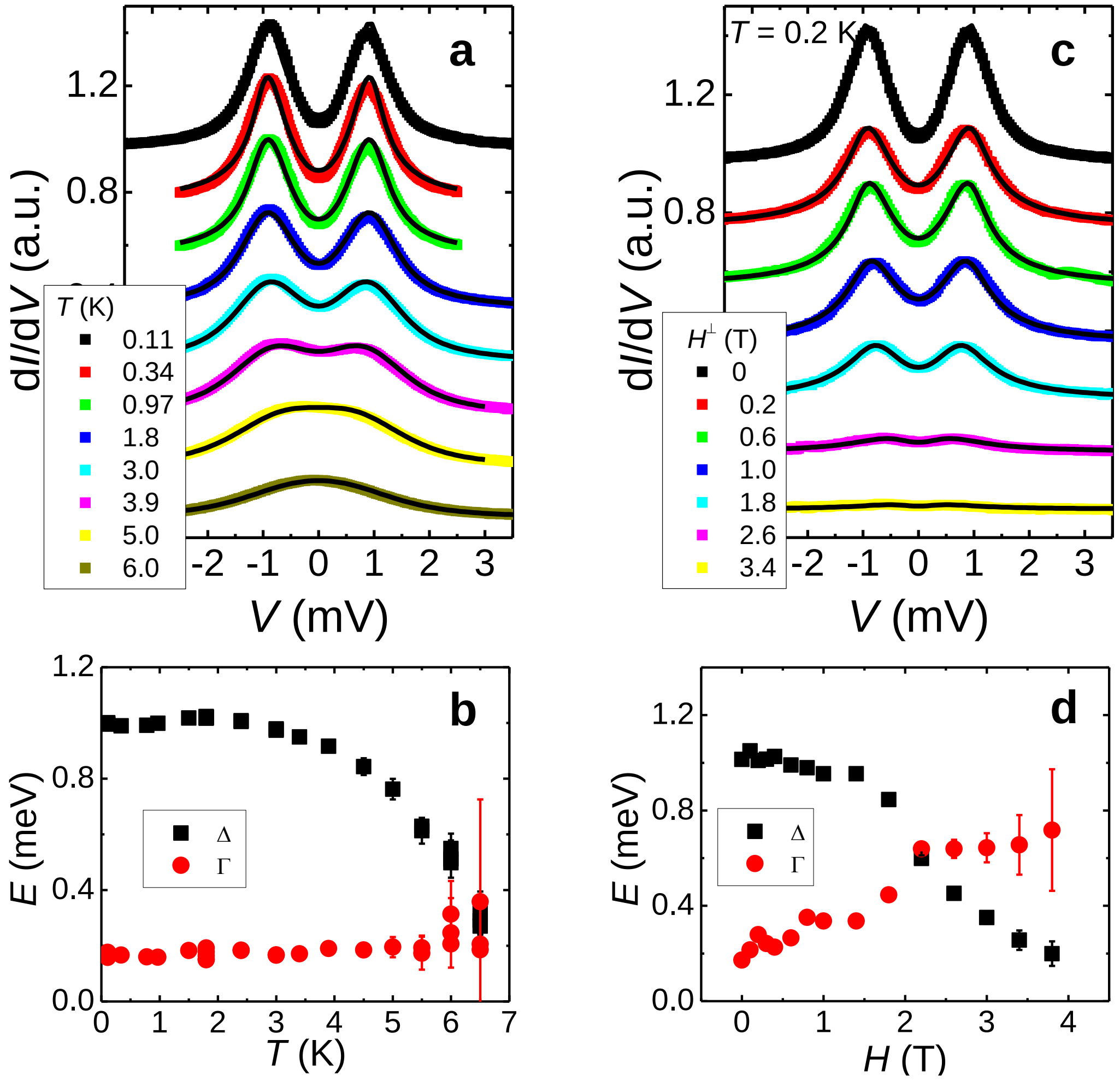

**Figure S1. Point contact conductance spectra of a $NbSe_2$/7 QL $Bi_2Se_3$ heterostructure**. **a,** Normalized conductance spectra of a point contact on the $NbSe_2$/7 QL $Bi_2Se_3$ sample at different temperatures. Curves are shifted vertically for clarity. The black lines are the BTK model fits to the experimental data. The contact normal resistance $R_n$ is 485 Ω. **b,** The temperature dependence of the superconducting energy gap Δ and quasiparticle lifetime broadening parameter Γ values from the fittings in Fig. S1**a**. **c,** Normalized conductance spectra of the point contact on the $NbSe_2$/7 QL $Bi_2Se_3$ sample at different magnetic fields. The sample temperature is 0.2 K. Curves are shifted vertically for clarity. The black lines are the BTK model fits to the experimental data. **d,** The magnetic field dependence of Δ and Γ values from the fittings in Fig. S1**c**.

## SI B. Point contact spectra of a $NbSe_2$/13 QL $Bi_2Se_3$ heterostructure

As we increase the $Bi_2Se_3$ film thickness, another conductance peak feature appears in the spectra and the spectra weight from the gap feature of $NbSe_2$ decreases. Figure S2**a** shows the conductance spectra of a point contact on a $NbSe_2$/13 QL $Bi_2Se_3$ heterostructure at different temperatures. At low temperatures, a conductance peak at low *V* bias ~ 0.3 mV appears besides the main $NbSe_2$ peak at ~ 1.0 mV. This gap feature is consistant with the proximity-induced bulk band energy gap. With increasing *T*, the two different peak features start to smear and become not resolvable above 3 K. Figure S2**b** shows the contact spectra at different magnetic fields at 0.1 K. The peak feature at ~ 0.3 mV is suppressed in small magnetic fields (see discussion in SI C). On the other hand, the $NbSe_2$ peak feature at ~ 1.0 mV remains unchanged in magnetic fields up to 20 mT. It disappears at ~ 4 T magnetic field, which is again consistent with the pure $NbSe_2$ $H_{c2}$.

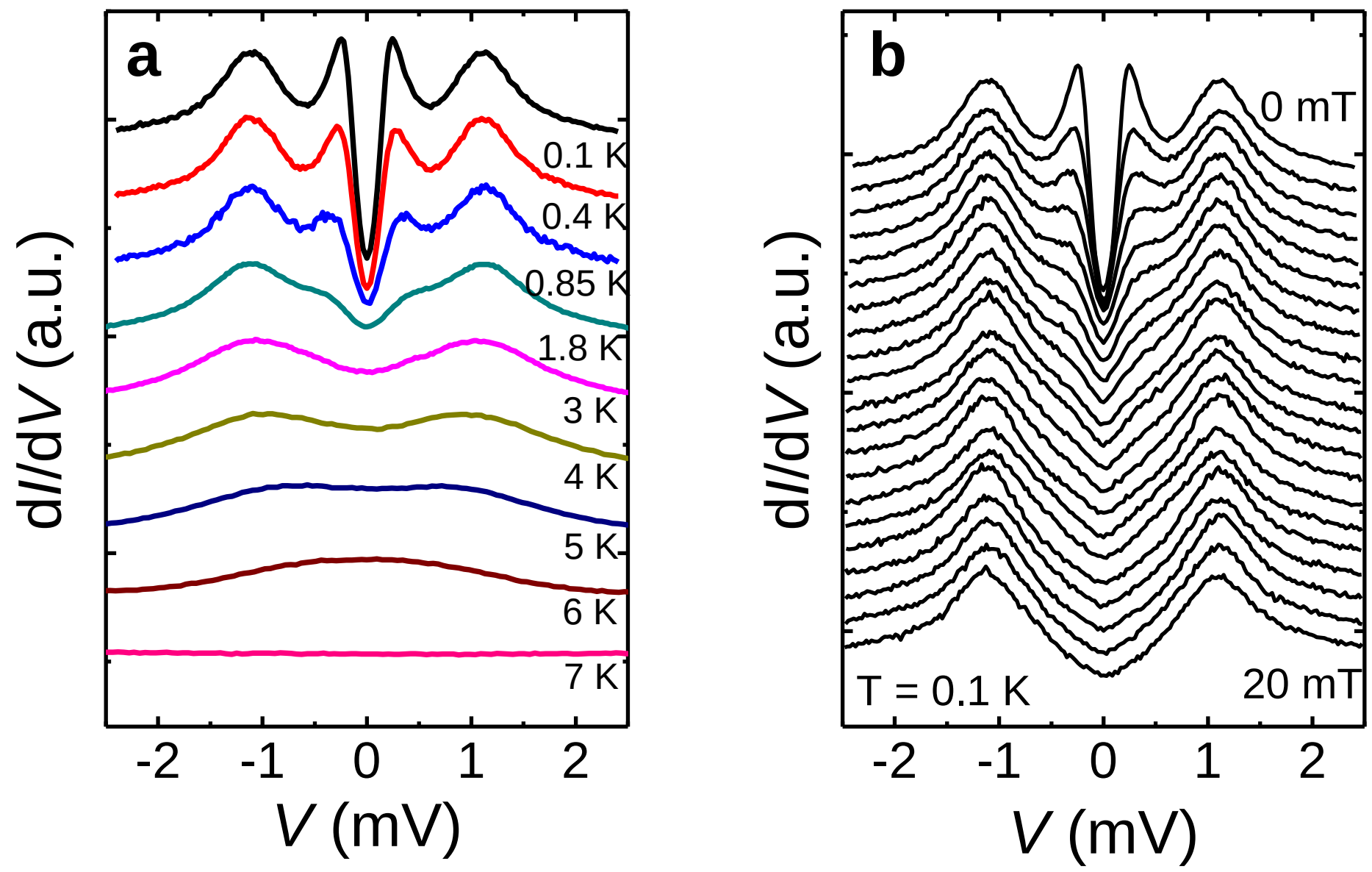

**Figure S2. Point contact conductance spectra of a $NbSe_2$/13 QL $Bi_2Se_3$ heterostructure. a,** Conductance spectra of a point contact junction on the $NbSe_2$/13 QL $Bi_2Se_3$ heterostructure at different temperatures. Curves are shifted vertically for clarity. **b,** The point contact spectra change with magnetic field from 0 to 20 mT at 0.1 K. Curves are shifted vertically. The contact normal resistance $R_n$ is 26 Ω.

## SI C. Magnetic field dependence of the proximity-induced bulk state gap in $Bi_2Se_3$

We studied the magnetic field dependence of the proximity-induced superconducting energy gap in the bulk of $Bi_2Se_3$. Figure S3**a** plots the point contact conductance spectra of the $NbSe_2$/16 QL $Bi_2Se_3$ heterostructure under different magnetic fields at 60 mK and the fittings using the BTK model. The energy gap $\Delta$ and broadening parameter $\Gamma$ from the fittings are ploted in Fig. S3**b** together with magnetic field dependence data at 1.8 K. In PCS studies under magnetic field, $\Gamma$ is often used to simulate the pair-breaking effect of a magnetic field in a first-order approximation.[3] We observed that the $\Gamma$ is proportional to $H$ at low magnetic fields. The $\Gamma/H$ ratio is ~ 3 meV/T for both 60 mK and 1.8 K, giving further support to the validity of using $\Gamma$ for field-induced broadening. The proximity-induced $Bi_2Se_3$ bulk state energy gap $\Delta$ decreases with increasing magnetic field. Although the conductance peak is nearly suppressed at $H$ = 0.3 T, there is a finite energy gap $\Delta$ from the BTK model fitting. It is likely that the broadening from the magnetic field-induced pair-breaking smears the $Bi_2Se_3$ bulk gap feature in the conductance spectra.

It is worth noting that our point contact spectroscopy probes a surface area much larger than the vortex core. The PCS signal is from averaging a large area containing vorices cores and superconducting areas between vortices. The PCS does not have the spatial resolution to resolve a single vortex.

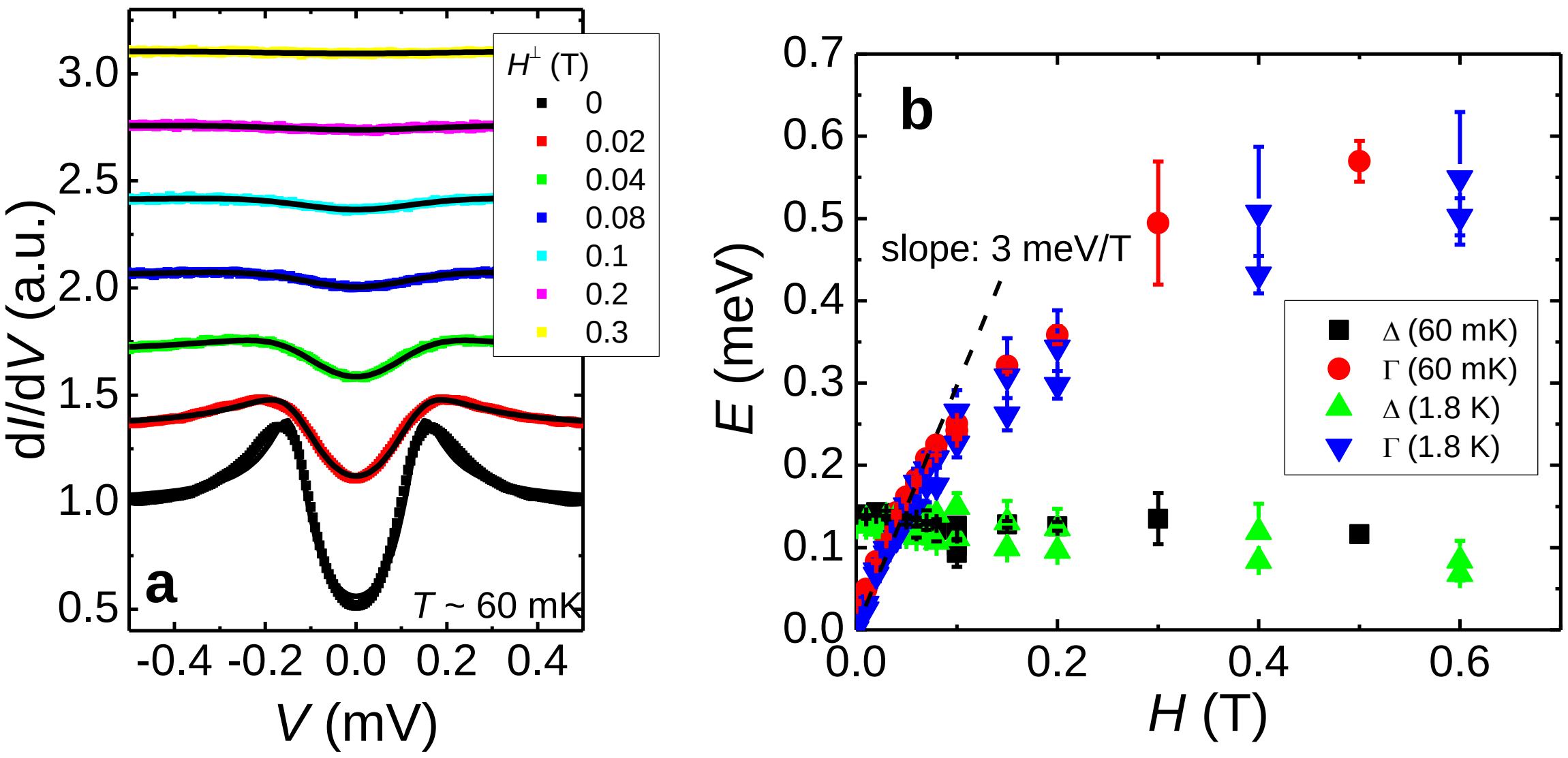

**Figure S3. Magnetic field dependence of the proximity-induced gap in $Bi_2Se_3$. a,** Pont contact conductance spectra of a $NbSe_2$/16 QL $Bi_2Se_3$ sample at 60 mK under different magnetic fields. Curves are shifted vertically for clarity. The black lines are BTK model fits to the experimental data. **b**, The magnetic field dependence of Δ and Γ values from BTK model fittings in Fig. S3**a** and fittings to point contact spectra at 1.8 K.

## SI D. Additional gap-like feature at low temperatures in point contact spectra of a $NbSe_2$/13 QL $Bi_2Se_3$ heterostructure

We also observed the additional gap-like feature in a 13 QL $Bi_2Se_3$ on $NbSe_2$ sample at low temperatures. Figure S4**a** shows the point contact spectra at different temperatures on a $NbSe_2$/13 QL $Bi_2Se_3$ heterostructure and the corresponding BTK model fittings. Similar to the $NbSe_2$/16 QL $Bi_2Se_3$ sample, the spectra of this 13 QL sample do not have the $NbSe_2$ peak feature so the one-gap standard BTK model is used to fit the conductance curves. The standard BTK model does not fit the experiemental data well at low temperatures; The fitted gap value decreases at low *T* (circled in Fig. S4**a** inset). Following the same method as used in the $NbSe_2$/16 QL $Bi_2Se_3$ sample analysis, a conductance cruve is calculated for $T$ = 0.4 K and plotted together with the experimental conductance spectrum in Fig. S4**b**. the condutance difference between the experiemental curve and the simulated curve shows a peak feature at ~ 210 μV (Fig. S4**b** inset). The ratio of this second gap feature to the main gap is ~ 0.7, close to the two induced gap ratio in the $NbSe_2$/16 QL $Bi_2Se_3$ sample at 0.2 K, indicating that the second gap feature in 13 QL and 16 QL samples are are from the same origin.

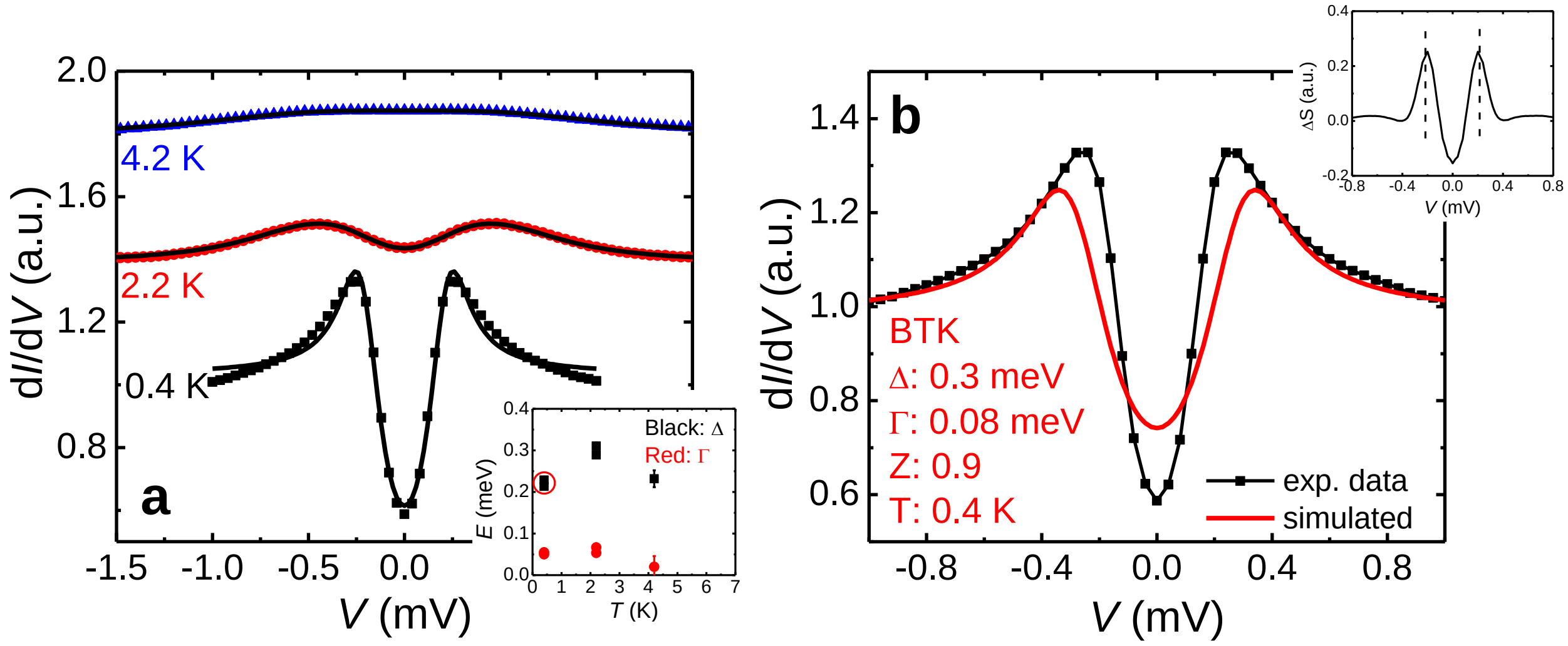

**Figure S4. Additional gap-like feature in point contact spectra of a $NbSe_2$/13 QL $Bi_2Se_3$ heterostructure at low temperatures. a,** Normalized point contact spectra of a $NbSe_2$/13 QL $Bi_2Se_3$ heterostructure at different temperatures. Curves are shifted vertically for clarity. The black lines are BTK model fits to the experimental data. The fitting parameters Δ and Γversus temperature are plotted in the bottom right inset. The contact normal resistance $R_n$ is 22 Ω. **b,** The point contact spectrum at 0.4 K (black) plotted together with a BTK model simulated spectrum using the parameters shown in the bottom left of the figure (red). The top right inset shows the conductance difference between the experimental data and the simulation as a function of bias $V$. The dash lines mark positions of peaks at ~ 210 μV.

## Reference

[1] Xu, S.-Y. *et al.* Momentum-space imaging of Cooper pairing in a half-Dirac-gas topological superconductor. *Nat. Phys.* **10,** 943–950 (2014).

[2] Blonder, G.E., Tinkham, M. & Klapwijk, T.M. Transition from metallic to tunneling regimes in superconducting microconstrictions: Excess current, charge imbalance, and supercurrent conversion. *Phys. Rev. B* **25,** 4515 (1982).

[3] Daghero, D. & Gonnelli, R.S. Probing multiband superconductivity by point-contact spectroscopy. *Supercond. Sci. Technol.* **23,** 43001 (2010).

[4] Gennes, P.G.D.  Superconductivity of Metals and Alloys. (W.A. Benjamin, New York, 1966).